\documentclass[twocolumn,showpacs,amsmath,amssymb,prl,nofootinbib]{revtex4}
\usepackage{graphicx}
\usepackage{dcolumn}
\usepackage{bm}

\begin{document}
\preprint{Draft-PRL}

\title{\boldmath Observation of the $Y(2175)$ in $J/\psi\rightarrow \eta\phi f_0(980)$}

\author{
M.~Ablikim$^{1}$,              J.~Z.~Bai$^{1}$,   Y.~Bai$^{1}$,
Y.~Ban$^{11}$, X.~Cai$^{1}$,                  H.~F.~Chen$^{16}$,
H.~S.~Chen$^{1}$,              H.~X.~Chen$^{1}$, J.~C.~Chen$^{1}$,
Jin~Chen$^{1}$
% ...
\\
\vspace{0.2cm}
(BES Collaboration)\\
\vspace{0.2cm} {\it
$^{1}$ Institute of High Energy Physics, Beijing 100049, People's Republic of China\\
$^{2}$ China Center for Advanced Science and Technology(CCAST),
Beijing 100080, People's Republic of China\\
$^{3}$ Guangxi Normal University, Guilin 541004, People's Republic of China\\
$^{4}$ Guangxi University, Nanning 530004, People's Republic of China\\
$^{5}$ Henan Normal University, Xinxiang 453002, People's Republic of China\\
$^{6}$ Huazhong Normal University, Wuhan 430079, People's Republic of China\\
$^{7}$ Hunan University, Changsha 410082, People's Republic of China\\
$^{8}$ Liaoning University, Shenyang 110036, People's Republic of China\\
$^{9}$ Nanjing Normal University, Nanjing 210097, People's Republic of China\\
$^{10}$ Nankai University, Tianjin 300071, People's Republic of China\\
$^{11}$ Peking University, Beijing 100871, People's Republic of China\\
$^{12}$ Shandong University, Jinan 250100, People's Republic of China\\
$^{13}$ Sichuan University, Chengdu 610064, People's Republic of China\\
$^{14}$ Tsinghua University, Beijing 100084, People's Republic of China\\
$^{15}$ University of Hawaii, Honolulu, HI 96822, USA\\
$^{16}$ University of Science and Technology of China, Hefei 230026,
People's Republic of China\\
$^{17}$ Wuhan University, Wuhan 430072, People's Republic of China\\
$^{18}$ Zhejiang University, Hangzhou 310028, People's Republic of China\\}
\vspace{0.2cm} {\it
$^{a}$ Current address: Johns Hopkins University, Baltimore, MD 21218, USA\\
$^{b}$ Current address: University of Oklahoma, Norman, Oklahoma 73019, USA\\
$^{c}$ Current address: DAPNIA/SPP Batiment 141, CEA Saclay, 91191, Gif sur Yvette Cedex, France\\
$^{d}$ Current address: Henan University of Science and Technology, Luoyang 471003, People's Republic of China\\
$^{e}$ Current address: CERN, CH-1211 Geneva 23, Switzerland\\
$^{f}$ Current address: University of Toronto, Toronto M5S 1A7, Canada\\
$^{g}$ Current address: Laboratoire de l'Acc{\'e}l{\'e}rateur
Lin{\'e}aire, Orsay, F-91898, France\\
$^{h}$ Current address: University of Colorado, Boulder, CO 80309, USA\\
$^{i}$ Current address: University of Michigan, Ann Arbor, MI 48109, USA\\}}

\vspace{0.4cm}

\date{Nov.28, 2007}

\begin{abstract}
The decays of $J/\psi\to \eta\phi f_0(980)~(\eta\to \gamma\gamma,
\phi \to K^+K^-, f_0(980)\to\pi^+\pi^-)$ are analyzed using a
sample of $5.8 \times 10^{7}$ $J/\psi$ events collected with the BESII
detector at the Beijing Electron-Positron Collider (BEPC). A structure
at around $2.18~$GeV/$c^2$ with about $5\sigma$ significance is
observed in the $\phi f_0(980)$ invariant mass spectrum. A fit
with a Breit-Wigner function gives the peak mass and width of
$m=2.186\pm 0.010~(stat)\pm 0.006~(syst)~$GeV/$c^2$ and
$\Gamma=0.065\pm 0.023~(stat)\pm 0.017~(syst)~$GeV/$c^2$,
respectively, which are consistent with those of $Y(2175)$,
observed by the BaBar collaboration in the initial-state
radiation (ISR) process $e^+e^-\to\gamma_{ISR}\phi f_0(980)$. The
production branching ratio is determined to be $Br(J/\psi\to\eta
Y(2175))\cdot Br(Y(2175)\to\phi f_0(980))\cdot
Br(f_0(980)\to\pi^+\pi^-)=(3.23\pm 0.75~(stat)\pm0.73~(syst))\times
10^{-4}$, assuming that the $Y(2175)$ is a $1^{--}$ state.
\end{abstract}
\pacs{13.25.Gv}
\maketitle

A new structure, denoted as $Y(2175)$ and with mass
$m=2.175\pm0.010\pm0.015$ GeV/$c^2$ and width
$\Gamma=58\pm16\pm20$ MeV/$c^2$, was observed by the BaBar
experiment in the $e^+e^-\to\gamma_{ISR}\phi f_0(980)$
initial-state radiation (ISR)
process~\cite{babary21752006, babary21752007}. This
observation stimulated some theoretical speculation that this
$J^{PC}=1^{--}$ state may be an $s$-quark version of the $Y(4260)$ since both of
them are produced in $e^+e^-$ annihilation and exhibit similar
decay patterns~\cite{babar4260}. There have been a number of
different interpretations proposed for the $Y(4260)$, including:
a $c \bar cg$ hybrid~\cite{y4260hybrid1, y4260hybrid2, y4260hybrid3};
a $4^3S_1$~$c\bar{c}$
state~\cite{y4260-4s}; a  $[cs]_S [\bar{c}\bar{s}]_S$ tetraquark
state~\cite{y4260-4quark}; or baryonium \cite{baryonium}. Likewise
a $Y(2175)$ has correspondingly been
interpreted as: a $s \bar s g$ hybrid~\cite{hybrid}; a $2^3D_1~s\bar s$
state~\cite{ssbar}; or a $s \bar ss\bar s$ tetraquark
state~\cite{tetraquark}. As of now, none of these interpretations
have either been established or ruled out by experiment.\\

\indent In this letter we report the observation of the $Y(2175)$ in
the decays of $J/\psi\to\eta\phi f_0(980)$, with $\eta\to
\gamma\gamma,~\phi \to K^+K^-,~f_0(980)\to\pi^+\pi^-$, using a
sample of $5.8 \times 10^{7}$ $J/\psi$ events collected with
the upgraded Beijing Spectrometer (BESII) detector at the Beijing
Electron-Positron Collider (BEPC).\\

\indent BESII is a large solid-angle magnetic spectrometer that is
described in detail in Ref.~\cite{BESII}. Charged particle momenta
are determined with a resolution of
$\sigma_p/p=1.78\%\sqrt{1+p^2}$ in a 40-layer cylindrical drift
chamber. Particle identification is accomplished using specific
ionization ($dE/dx$) measurements in the main drift chamber (MDC)
and time-of-flight (TOF) measurements in a barrel-like array of 48
scintillation counters. The $dE/dx$ resolution is
$\sigma_{dE/dx}=8.0\%$, and the TOF resolution is
$\sigma_{TOF}=180$ ps for Bhabha tracks. Outside of the
time-of-flight counters is a 12-radiation-length barrel shower
counter (BSC) comprised of gas tubes interleaved with lead sheets.
The BSC measures the energies and directions of photons with
resolutions of $\sigma_E/E\simeq 21\%/\sqrt{E(\mbox{GeV})}$,
$\sigma_{\phi} = 7.9$ mrad, and $\sigma_{z}$ = 2.3 cm. The iron
flux return of the magnet is instrumented with three double layers
of counters that are used to identify muons.\\

\indent In this analysis, a GEANT3-based Monte Carlo (MC) package
with detailed consideration of the detector performance is used.
The consistency between data and MC has been validated using
many high purity physics channels~\cite{simbes}. For $J/\psi \to
\eta Y(2175)(Y(2175) \to \phi f_0(980), f_0(980) \to \pi^+\pi^-)$,
a Monte-Carlo generator that assumes the $Y(2175)$ quantum numbers
to be $J^{PC} =1^{--}$ and considers the angular distributions
for $1^{--} \to 0^{-+} + 1^{--}$; $1^{--}\to 1^{--}+ 0^{++}$
is used to determine the detection  efficiency.\\

\begin{figure}[htbp]
  \centering
\includegraphics[width=7.5cm,height=14cm]{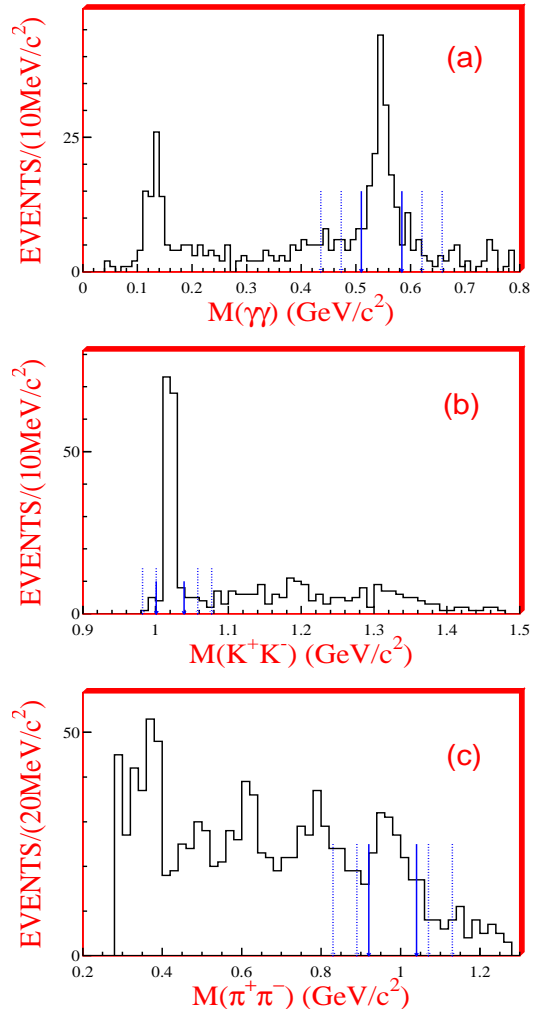}
  \caption{\footnotesize (a) The $\gamma\gamma$ invariant mass
  spectrum. (b) The $K^+K^-$ invariant mass spectrum.
  (c) The $\pi^+\pi^-$ invariant mass spectrum.
  The solid arrows in each plot show the cuts imposed for $\eta$,
  $\phi$ and $f_0$ selection. The dashed arrows show the sideband regions used
  to estimate background levels.}
           \label{draft-mass}
\end{figure}

\begin{figure}[htbp]
  \centering
\includegraphics[width=6cm,height=4cm]{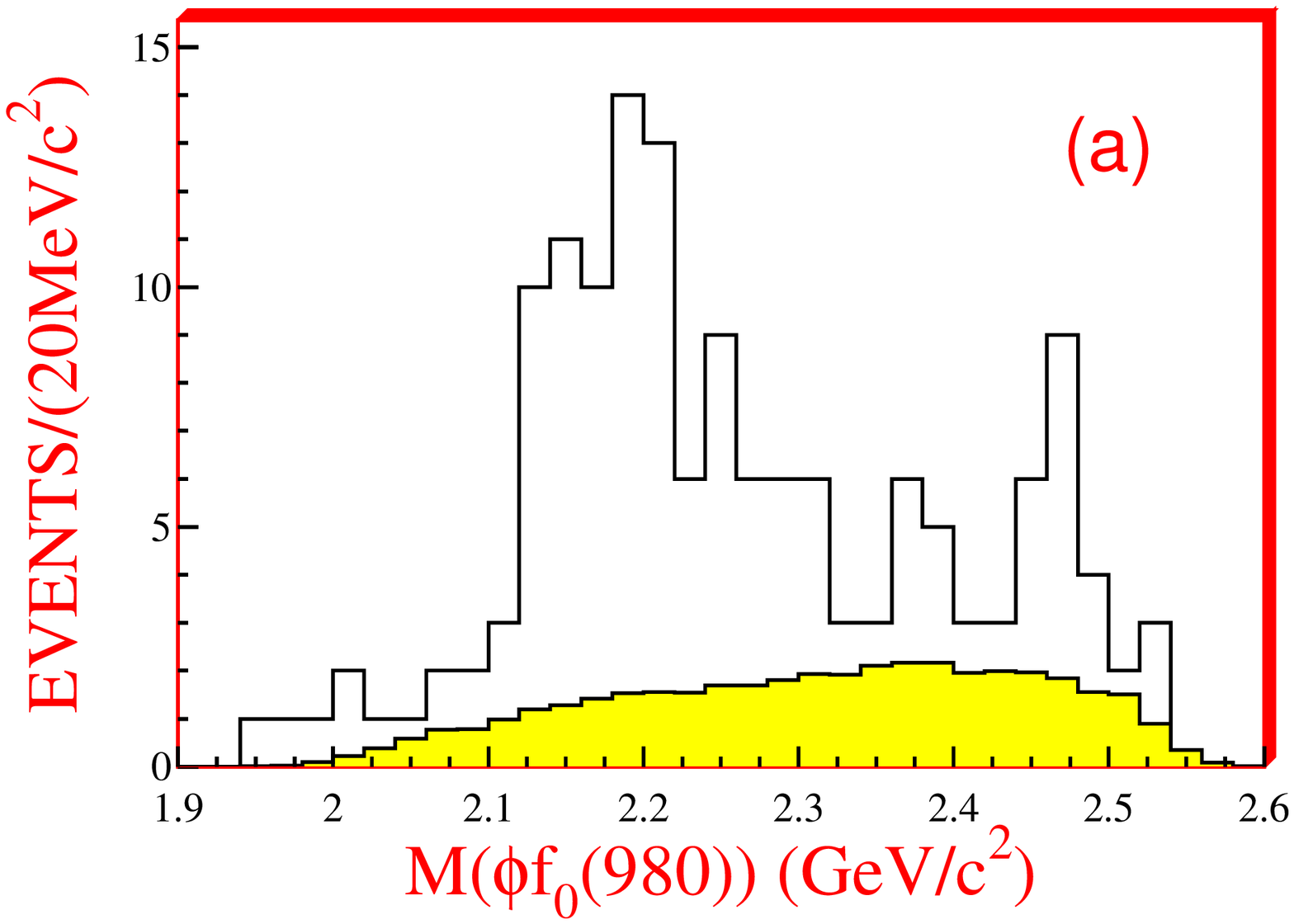}
\includegraphics[width=6cm,height=6cm]{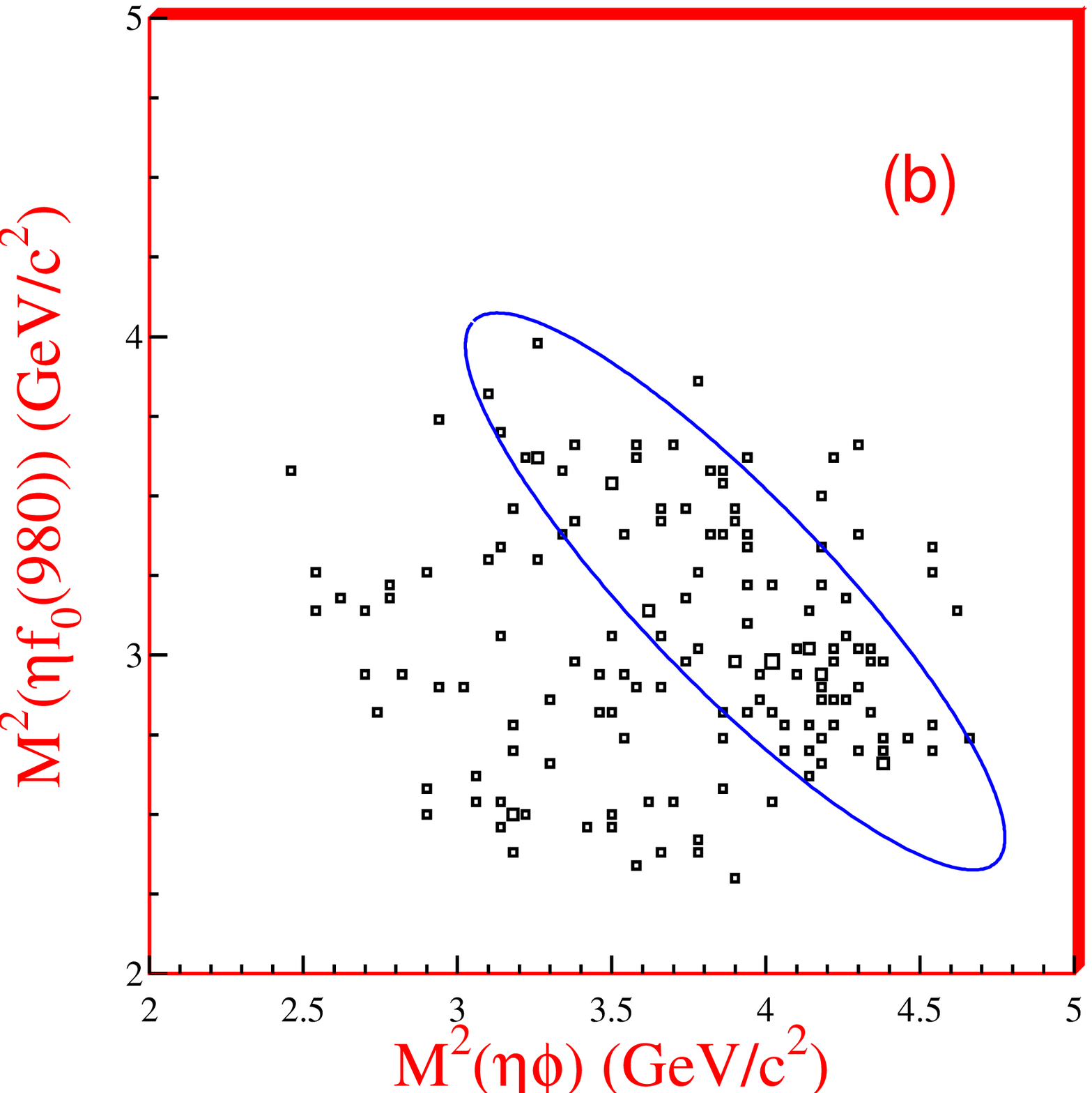}
  \caption{\footnotesize (a) The $\phi f_0(980)$ invariant
  mass spectrum. The open histogram is data and the shaded
  histogram is $J/\psi~\to~\eta\phi f_0(980)$ phase-space MC
  events (with arbitrary normalization). (b) The Dalitz plot of
  $m^2_{\eta f_0(980)}$ versus $m^2_{\eta\phi}$. The ellipse
  shows the resonance band in $\phi f_0(980)$
  invariant mass spectrum.
  }
           \label{draft-mass2}
\end{figure}

\indent For a candidate event, we require four good charged tracks
with zero net charge. A good charged track is  one that can be well fitted
to a helix within the polar angle region $|\cos \theta|<0.8$
and has a transverse momentum larger than $70$~MeV/$c$. For each
charged track, the TOF and $dE/dx$ information are combined to form
particle identification confidence levels for the $\pi$, $K$ and $p$
hypotheses; the particle type with the highest confidence level is
assigned to each track. The four charged tracks are required to
consist of an unambiguously identified $K^+ K^-\pi^+\pi^-$
combination. Candidate photons are required to have an energy
deposited in the BSC that is greater than 60 MeV and to be isolated
from charged tracks by more than $5^{\circ}$; at least two photons
are required. A four-constraint (4C) energy-momentum conservation
kinematic fit is performed to the $K^+ K^-\pi^+\pi^-\gamma\gamma$
hypothesis and the $\chi^{2}_{4C}$ is required to be less than 15.
For events with more than two selected photons, the combination with
the smallest $\chi^{2}$ is chosen. An $\eta$ signal is evident in
the $\gamma\gamma$ invariant mass spectrum
(Fig.~\ref{draft-mass}(a)); $\eta\to\gamma\gamma$ candidates are
defined as $\gamma$-pairs with $|M_{\gamma\gamma}-0.547|<0.037$
GeV/$c^2$. A $\phi$ signal is distinct in the $K^+K^-$ invariant
mass spectrum (Fig.~\ref{draft-mass}(b)), and for these candidates, we
require $|m_{K^+K^-}-1.02|<0.019$GeV/$c^2$. In the $\pi^+\pi^-$
invariant mass spectrum, candidate $f_0(980)$ mesons are defined by
$|m_{\pi^+\pi^-}-0.980|<0.060$GeV/$c^2$ (Fig.~\ref{draft-mass}(c)).
The $\phi f_0(980)$ invariant mass spectrum for the selected events
is shown in Fig.~\ref{draft-mass2}(a),
where a clear enhancement is seen around $2.18~$ GeV/$c^2$.\\
\indent The Dalitz plot of $m^2_{\eta f_0(980)}$ versus
$m^2_{\eta\phi}$ for the selected events is shown in
Fig.~\ref{draft-mass2}(b), where a diagonal band can be seen. This
band corresponds to the structure observed around 2.18 GeV/$c^2$ in
the $\phi f_0(980)$
invariant mass spectrum shown in Fig.~\ref{draft-mass2}(a).\\

\begin{figure}[htbp]
  \centering
\includegraphics[width=6.5cm,height=5cm]{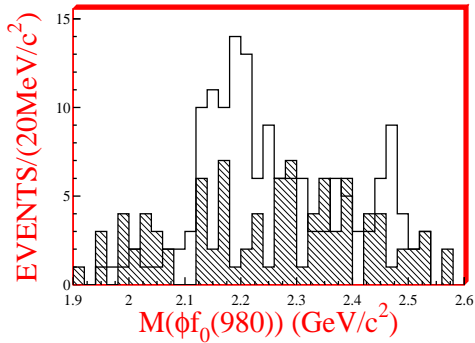}
  \caption{\footnotesize The $\phi f_0(980)$ invariant mass spectrum.
  The open histogram is data and the shaded
  histogram shows the sideband-determined background.}
           \label{draft-sideband}
\end{figure}

\indent To clarify the origin of the observed structure, we have
made extensive studies of potential background processes using both
data and MC. Non-$\eta$ or non-$f_0(980)$  processes are studied
with $\eta$-$f_0(980)$ mass sideband
events~(0.074 GeV/$c^2<|M_{\gamma\gamma}-0.547|<0.111$GeV/$c^2$ or
0.090 GeV/$c^2<|m_{\pi^+\pi^-}-0.980|<0.150$GeV/$c^2$). Non-$\phi$
processes are studied with $\phi$ mass sideband events
(0.038 GeV/$c^2<(m_{K^+K^-}-1.02)<0.057$GeV/$c^2$ or
-0.038 GeV/$c^2<(m_{K^+K^-}-1.02)<-0.019$GeV/$c^2$). The scaled $M_{\pi^+\pi^-K^+K^-}$
distribution for the summed total of
sideband events (minus double
counting) are shown as a shaded histogram in Fig.~\ref{draft-sideband}. No structure
around $2.18~$GeV/$c^2$ is evident. In addition, we also checked for
possible backgrounds from various $J/\psi$ decays using
Monte-Carlo simulation, and no evidence of a
structure at $2.18 ~$GeV/$c^2$ is observed.\\

\indent We fit the $\phi f_0(980)$ invariant mass spectrum (see
Fig.~\ref{draft-mass2}(a)) and the total sidebands (see
Fig.~\ref{draft-sideband}) simultaneously. The procedure is as
follows: First we fit the sideband distribution with a 3rd-order
polynomial. Next we use the polynomial shape as the background
function for both the $\phi f_0(980)$ invariant mass spectrum
histogram and the total sideband histogram, and the signal and
background normalizations are allowed to float. In this fit, the
normalization for the background polynomial is constrained to be the
same for both the signal and sideband histograms. We use a
constant-width Breit-Wigner (BW) convolved with a Gaussian mass
resolution function (with $\sigma$ = $12$MeV/$c^2$) to represent the
$Y(2175)$ signal. The mass and width obtained from the fit (shown as
smooth curves in Fig.~\ref{draft-fit}) are $m=2.186\pm
0.010~(stat)~$GeV/$c^2$ and $\Gamma=0.065\pm
0.023~(stat)~$GeV/$c^2$. The fit yields $52\pm12$ signal events and
$-2lnL$~($L$ is the likelihood value of the fit)~$=~78.6$. A fit to
the mass spectrum without a BW signal function returns
$-2lnL~=~116.0$. The change in $-2lnL$ with a change of degrees of
freedom $=$ 3 corresponds to a statistical
significance of $5.5~\sigma$ for the signal.\\
\indent Using the MC-determined selection efficiency of $1.44\%$, we
find the product branching ratio to be:

\begin{figure}[htbp]
  \centering
\includegraphics[width=6cm,height=4cm]{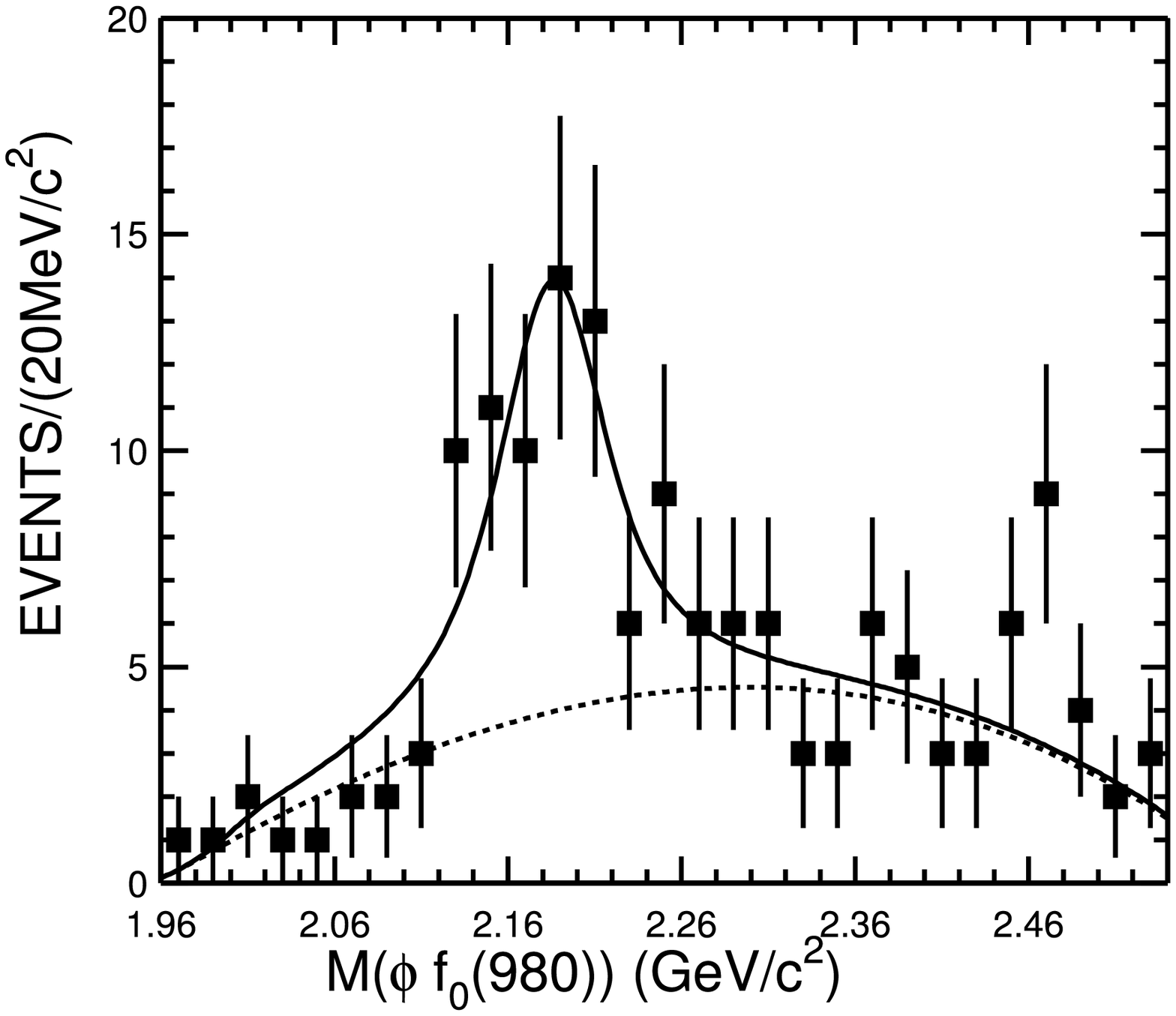}
\includegraphics[width=6cm,height=4cm]{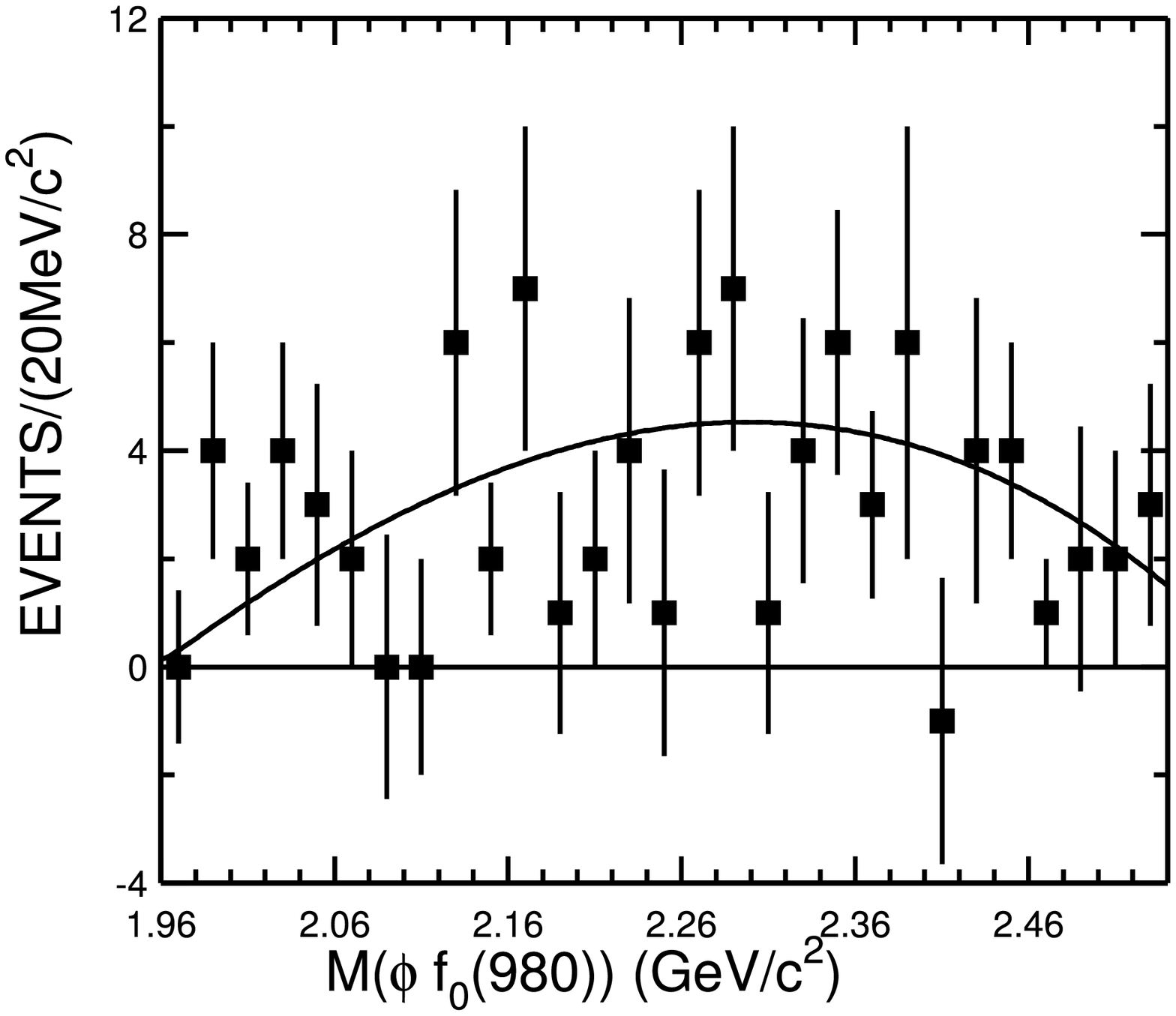}
  \caption{\footnotesize The top panel shows the fit (solid curve)
to the data (points with error bars);~the dashed curve indicates
the background function. The bottom panel shows the simultaneous
fit to the sideband events (points with error bars) with the same
background function. The background normalizations for the two
plots are constrained to be equal.}
           \label{draft-fit}
\end{figure}

\begin{center}
$Br(J/\psi \to \eta Y(2175))\cdot Br(Y(2175)\to \phi
  f_0(980))Br(f_0(980)\to \pi^+\pi^-)=(3.23\pm 0.75)\times
  10^{-4}$.
\end{center}

Fits that use different treatments for the background are also tried. If
the background is fitted as a 3rd-order polynomial with all
parameters allowed to float, the signal yield is $61 \pm 14$
events, with mass and width of $m=2.182\pm
0.010~(stat)~$GeV/$c^2$ and $\Gamma=0.073\pm
0.024~(stat)~$GeV/$c^2$, respectively. The statistical significance
is $4.9 ~\sigma$. If the background shape is fixed to the shape of
phase space, the fit yields $57 \pm 13$ signal events, with a
statistical significance of $5.3 ~\sigma$. The mass and width
obtained are $m=2.182\pm 0.009~(stat)~$GeV/$c^2$ and
$\Gamma=0.069\pm 0.022~(stat)~$GeV/$c^2$. For all of the background
shapes considered, the fitted masses and widths of the signal are
consistent with each other. We take the results with the background
shape fixed to the sideband shape as the central values.

We determine the systematic uncertainties of the mass and width measurements by
varying the functional form used to represent the background, the
fitting range of the invariant mass spectrum, the bin width of the
invariant mass spectrum, allowing the sideband and signal
background normalizations to differ, and including possible fitting biases.
The latter are estimated from the differences between the input
and output mass and width values from a MC study. Adding each
contribution in quadrature, the total
systematic errors on the mass and width are 6 MeV/$c^2$ and 17
MeV/$c^2$, respectively. The systematic error on the branching
ratio measurement comes mainly from the uncertainties in the MDC
simulation (including systematic uncertainties of the tracking
efficiency and the kinematic fits), the photon detection
efficiency, the particle identification efficiency, the $\eta$
decay branching ratio to $\gamma\gamma$ and the $\phi$ decay
branching ratio to $K^+K^-$, the background function, the fitting
range of the invariant mass spectrum, the bin width of the
invariant mass spectrum, the fitting method and the total number
of $J/\psi $ events~\cite{ssfang}. Adding all contributions in
quadrature gives a total systematic error on
the product branching ratio of
$22.7\%$.\\

\indent We studied the small peak near 2.47 GeV/$c^2$ in the
$\phi f_0(980)$ invariant mass spectrum (see
Fig.~\ref{draft-mass2}(a)), which was also noted by BaBar~\cite{babary21752007}.
A fit was made to the $\phi f_0(980)$
invariant mass spectrum using two non-interfering Breit-Wigner
functions with mass and width of the second peak fixed
to the BaBar fitted results: $2.47~$GeV/$c^2$ and
$0.077~$GeV/$c^2$~\cite{babary21752007}, respectively. The fit results
indicate a significance for the first peak of $5.8 ~\sigma$, with
a mass and width of $m=2.186\pm 0.010~(stat)~$GeV/$c^2$ and
$\Gamma=0.065\pm 0.022~(stat)~$GeV/$c^2$, respectively. The
statistical significance of the second peak is only $2.5~\sigma$.\\

\indent In summary, the $J/\psi \to \eta \phi f_0(980)$ decay
process with $\eta \to \gamma \gamma$, $\phi \to K^+ K^-$, and
$f_0(980) \to \pi^+\pi^-$ has been analyzed. A structure, the $Y(2175)$,
is observed with about $5\sigma$ significance in the $\phi f_0(980)$
invariant mass spectrum. From a fit with a Breit-Wigner function,
the mass is determined to be $M=2.186\pm 0.010~(stat)\pm
0.006~(syst)~$GeV/$c^2$ , the width is $\Gamma=0.065\pm
0.023~(stat)\pm 0.017~(syst)~$GeV/$c^2$ and the product branching
ratio is $Br(J/\psi \to \eta Y(2175))\cdot Br(Y(2175)\to \phi
f_0(980))\cdot Br(f_0(980)\to\pi^+\pi^-)=(3.23\pm 0.75~(stat)\pm
0.73~(syst))\times 10^{-4}$. The mass and width are consistent with
BaBar's results. The identification of the precise nature of the
$Y(2175)$ requires measurements of additional decay channels~\cite{hybrid, ssbar}.
This is the subject of the work that is currently in progress.\\

\indent The BES collaboration thanks the staff of BEPC and
computing center for their hard efforts. This work is supported in
part by the National Natural Science Foundation of China under
contracts Nos. 10491300, 10225524, 10225525, 10425523, 10625524,
10521003, the Chinese Academy of Sciences under contract No. KJ
95T-03, the 100 Talents Program of CAS under Contract Nos. U-11,
U-24, U-25, and the Knowledge Innovation Project of CAS under
Contract Nos. U-602, U-34 (IHEP), the National Natural Science
Foundation of China under Contract No. 10225522 (Tsinghua
University), and the Department of Energy under Contract
No.DE-FG02-04ER41291 (U. Hawaii).

\end{document}